\documentclass[10pt, conference, letterpaper]{IEEEtran}
\usepackage[utf8]{inputenc}

\usepackage[T1]{fontenc}
\usepackage{lineno}
\usepackage{graphicx}
\usepackage{subcaption}
\usepackage{algorithm}
\usepackage{algpseudocode}
\usepackage{amsmath}
\usepackage{enumitem}
\usepackage{multirow}
\usepackage{url}
\usepackage{booktabs}

\usepackage{xcolor}        
\definecolor{kwblue}{RGB}{10,102,194} 
\definecolor{bggray}{RGB}{245,245,245}  
\definecolor{strorange}{RGB}{204,102,0} 
\definecolor{cmtgreen}{RGB}{0,128,0} 
\usepackage{listings}      

\lstdefinestyle{bashstyle}{
    backgroundcolor=\color{bggray},          
    basicstyle=\ttfamily\small,              
    keywordstyle=\color{kwblue}\bfseries,    
    stringstyle=\color{strorange},           
    commentstyle=\color{cmtgreen}\itshape,   
    frame=single,                             
    rulecolor=\color{gray},                  
    numbers=left,                             
    numberstyle=\tiny\color{gray},           
    numbersep=5pt,                            
    breaklines=true,                          
    showstringspaces=false                    
}

\title{MH-1M: A 1.34 Million-Sample Comprehensive Multi-Feature Android Malware Dataset for Machine Learning, Deep Learning, Large Language Models, and Threat Intelligence Research}

\author{
Hendrio Bragança$^{1}$, Diego Kreutz$^{2}$, Vanderson Rocha$^{1}$, Joner Assolin$^{1}$, and Eduardo Feitosa$^{1}$%
\thanks{$^{1}$Federal University of Amazonas (UFAM), Institute of Computing, Manaus, 69077-000, Amazonas, Brazil.}%
\thanks{$^{2}$Federal University of Pampa (UNIPAMPA), AI Horizon Labs, Alegrete, RS, 97546-550, Brazil.}%
\thanks{*Corresponding author: Hendrio Bragança (hendrio.luis@icomp.ufam.edu.br)}%
}

\begin{document}

\maketitle

\begin{abstract}
We present MH-1M, one of the most comprehensive and up-to-date datasets for advanced Android malware research.
The dataset comprises 1,340,515 applications, encompassing a wide range of features and extensive metadata.
To ensure accurate malware classification, we employ the VirusTotal API, integrating multiple detection engines for comprehensive and reliable assessment.
Our GitHub, Figshare, and Harvard Dataverse repositories provide open access to the processed dataset and its extensive supplementary metadata, totaling more than 400 GB of data and including the outputs of the feature extraction pipeline as well as the corresponding VirusTotal reports.
Our findings underscore the MH-1M dataset's invaluable role in understanding the evolving landscape of malware.

\end{abstract}



\section*{Background \& Summary}

The pervasive spread of Android malware poses a significant challenge for cybersecurity research. 
This challenge stems mainly from the open-source nature and affordability of Android platforms, which grant users access to a large market of free applications. 
At the same time, malware continually evolves, adapting its tactics to execute more sophisticated and frequent attacks. 
Such attacks often result in data destruction, information theft, and several other cybercrimes \cite{aboaoja2022malware, miranda2022debiasing, kumar2023understanding}.

Machine learning (ML) algorithms have been widely used to uncover malware and have demonstrated remarkable effectiveness in detection systems, leveraging their discriminative capabilities to identify new variants of malicious applications \cite{scalas2021malware,10734108,10.1145/3638240}. 
To mitigate these risks, researchers have developed a variety of methods for detecting Android malware, establishing machine learning as a central focus of contemporary mobile security research \cite{zakeya2022probing,10504267,10.1145/3691621.3694948}.

However, the effectiveness of ML models is highly dependent on the quality of the datasets used for training. 
Many existing datasets suffer from limitations such as outdated data, inadequate representation, and a limited number of samples and features, making them unsuitable for modern malware detection \cite{botacin2021challenges,miranda2022debiasing,10529242,BHARDWAJ2024103769}. 
These issues raise concerns about the reliability of reported performance metrics and can potentially lead to misleading conclusions \cite{miranda2022debiasing}.
A growing body of research in Android malware detection strongly supports the notion that increasing the number of discriminative features can significantly improve classification performance \cite{bragancca2023android,10.1145/3678890.3678893,Hasan2025MalwareDetection}.

We present in Table \ref{tab:datasets} an overview of widely used Android malware datasets from recent years.
Datasets such as AndroCrawl \cite{sisto2012androcrawl}, ADROIT \cite{7849904}, Android Permissions \cite{10.1145/2295136.2295141}, DefenseDroid\footnote{\url{https://github.com/DefenseDroid/DefenseDroid?tab=readme-ov-file}}, DREBIN-215 \cite{Yerima2018Android}, KronoDroid \cite{guerra2021kronodroid}, and MH-100K \cite{bragancca2023android} are frequently referenced in the literature but exhibit several notable limitations.
First, they are generally limited in scale, restricting the diversity of samples available for training and evaluation.
Second, most datasets provide a narrow feature scope, often focusing on a single aspect, such as permissions or API calls.
Third, many of them include a constrained and, in some cases, outdated malware taxonomy with inconsistent or poorly documented labeling criteria.

\begin{table*}[p]
\caption{Summarization of most common Android Malware datasets.}
\label{tab:datasets}
\centering
\begin{tabular}{@{}lclclc@{}}
\toprule
\multicolumn{1}{c}{\multirow{2}{*}{Dataset}} & \multicolumn{2}{c}{Features} & \multicolumn{3}{c}{Samples} \\ \cmidrule(l){2-6} 
\multicolumn{1}{c}{} & \textbf{N. Features} & \textbf{Feature Type} & \textbf{Malwares} & \textbf{Benign} & \textbf{Total} \\ \midrule
AndroCrawl & 81 & \begin{tabular}[c]{@{}l@{}}API Calls (24) \\ Intents (8)\\ Permissions (49)\end{tabular} & 10,170 & 86,562 & 96,732 \\\hline
ADROIT & 166 & Permissions & 3,418 & 8,058 & 11,476 \\\hline
Android Permissions & 183 & Permissions & 20,000 & 9,999 & 29,999 \\\hline
DefenseDroid & 2938 & \begin{tabular}[c]{@{}l@{}}Permissions (1490)\\ Intents (1448)\end{tabular} & 6,000 & 5,975 & 11,975 \\\hline
DREBIN-215 & 215 & \begin{tabular}[c]{@{}l@{}}API Calls (73)\\ Permissions (113)\\ System Commands (6)\\ Intents (23)\end{tabular} & 5,555 & 9,476 & 15,036 \\\hline
\begin{tabular}[c]{@{}l@{}}KronoDroid Disp. Real\end{tabular} & 246 & \begin{tabular}[c]{@{}l@{}}Permissions (146)\\ System Calls (100)\end{tabular} & 41,382 & 36,755 & 78,137 \\\hline
MH-100K & 24833 & \begin{tabular}[c]{@{}l@{}}API Calls (24417) \\ Intents (250)\\ Permissions (166)\end{tabular} & 9,800 & 92,134 & 101,934 \\\hline
MH-1M & 23247 & \begin{tabular}[c]{@{}l@{}}API Calls (22394)\\ Intents (407)\\ Permissions (214)\\ Opcodes (232)\end{tabular} & 119,094 & 1,221,421 & 1,340,515 \\
\bottomrule
\end{tabular}
\end{table*}


Table \ref{tab:model_performance} demonstrates that the wide diversity of Android malware behaviors, often represented by non-overlapping feature sets, highlights the importance of adopting multidimensional feature representations for comprehensive analysis and effective detection \cite{braganca2023xai,Nazim2025MultimodalMalware,Wang2024ExploringMalware}.
For instance, the Drebin-215 dataset \cite{Yerima2018Android}, released in 2018, is derived from the original Drebin dataset developed in 2012. Consequently, models trained on it rely on outdated samples that no longer reflect the current Android malware landscape.
Likewise, the CICInvesAndMal2019 dataset \cite{Taheri2019Investigation}, although reported to include data from 2019, still incorporates features extracted from obsolete Android API versions (2016 and earlier), limiting its relevance for modern threat detection.

\begin{table*}[p]
\centering
\caption{Classification performance using holdout methodology for malware detection models on different datasets.}
\begin{tabular}{@{}lcccccc@{}}
\toprule
Dataset & Class & Precision & Recall & F1-score & Accuracy & Macro-F1 \\ \midrule
\multirow{2}{*}{Drebin} & Benign & 0.99 & 0.99 & 0.99 & \multirow{2}{*}{0.99} & \multirow{2}{*}{0.99} \\
 & Malware & 0.99 & 0.99 & 0.99 &  &  \\ \midrule
\multirow{2}{*}{AndroCrawl} & Benign & 0.99 & 0.99 & 0.99 & \multirow{2}{*}{0.99} & \multirow{2}{*}{0.97} \\
 & Malware & 0.94 & 0.94 & 0.94 &  &  \\ \midrule
\multirow{2}{*}{KronoDroid} & Benign & 0.96 & 0.98 & 0.97 & \multirow{2}{*}{0.97} & \multirow{2}{*}{0.97} \\
 & Malware & 0.98 & 0.97 & 0.97 &  &  \\ \midrule
\multirow{2}{*}{AndroidPermissions} & Benign & 0.55 & 0.13 & 0.21 & \multirow{2}{*}{0.67} & \multirow{2}{*}{0.50} \\
 & Malware & 0.68 & 0.94 & 0.79 &  &  \\ \midrule
\multirow{2}{*}{Adroit} & Benign & 0.91 & 0.97 & 0.94 & \multirow{2}{*}{0.91} & \multirow{2}{*}{0.89} \\
 & Malware & 0.92 & 0.78 & 0.85 &  &  \\ \midrule
\multirow{2}{*}{DefenseDroid} & Benign & 0.91 & 0.93 & 0.92 & \multirow{2}{*}{0.92} & \multirow{2}{*}{0.92} \\
 & Malware & 0.93 & 0.90 & 0.92 &  &  \\ \midrule
\multirow{2}{*}{MH-100K} & Benign & 0.99 & 0.99 & 0.99 & \multirow{2}{*}{0.98} & \multirow{2}{*}{0.94} \\
 & Malware & 0.87 & 0.90 & 0.89 &  &  \\ \bottomrule
\end{tabular}%
\label{tab:model_performance}
\end{table*}

While Android malware samples may share common behaviors and exploitation techniques, reflected in overlapping feature categories, the specific features used for detection can differ considerably across datasets. This variability underscores the evolving and complex nature of Android malware, which continuously adapts to new defense mechanisms. Moreover, recent studies indicate a broader challenge in the field of artificial intelligence, showing that up to 80\% of AI projects fail primarily due to the lack of representative and high-quality datasets \cite{schmelzer2022the, ai2023top}. Inadequate data quality remains one of the most frequently cited obstacles to the development of robust, realistic, and generalizable machine learning models \cite{10845793, NASER2025103122}.

To address this critical issue in AI research and to advance the field of Android malware detection, we present the MH-1M Android Malware dataset, one of the most extensive and comprehensive datasets ever compiled for this purpose. The dataset contains 1,340,515 Android application packages (APKs) and 22,810 extracted attributes, all obtained from the AndroZoo repository\footnote{\url{https://androzoo.uni.lu}}, covering a period of 14 years, from 2010 to 2024. For comparison, one of the largest previously available datasets, Drebin \cite{arp2014drebin}, was released in 2014 and included approximately 1 million samples. More recently, the MH-100K dataset \cite{bragancca2023android}, published in late 2023, provided around 100,000 samples but lacked the extensive metadata and temporal diversity present in MH-1M.

MH-1M not only surpasses Drebin in scale but also incorporates updated and comprehensive information that is essential for advancing modern malware detection methodologies. In addition, MH-1M provides extensive metadata generated during the extraction and analysis processes, a feature not present in previous datasets. This metadata includes detailed information about the feature extraction procedures and corresponding analysis results, thereby improving transparency and reproducibility in research on Android malware. Furthermore, the inclusion of both structured and unstructured metadata expands the potential for applying deep learning and large language model (LLM) approaches to more sophisticated and context-aware malware detection.

It is safe to say that MH-1M provides an unparalleled collection of metadata from APKs, offering insights into the evolution of malicious software that spans over a decade. 
It includes detailed Android features such as 22,394 API calls, 407 intents, 232 opcodes, and 214 permissions. 
These are supplemented by extensive metadata, including SHA-256 hashes, file names, package names, compilation APIs, VirusTotal reports, and additional contextual attributes that provide a rich foundation for advanced research on Android malware detection. In total, the dataset offers more than 400 GB of valuable data, available in compressed format at the Harvard Dataverse repository \cite{Braganca2025MH1M}, representing the largest and most comprehensive collection ever compiled to advance research and development in Android malware detection.

We used the VirusTotal Web API to evaluate the threat level of each Android sample based on the results of multiple detection engines. The VirusTotal API offers a comprehensive view of the maliciousness of an application by aggregating the outcomes of numerous antivirus and security tools. Among the available indicators, the detection count is commonly employed as a proxy for label confidence in malware analysis. For instance, a file detected by only one or two engines may represent a false positive, whereas a file flagged by a higher number of engines (e.g., eight or more) is generally considered genuinely malicious. In academic and machine learning studies, this detection count is frequently used to establish labeling thresholds and define the ground truth for malware datasets.

In our study, we evaluated seven widely used Android malware datasets (Table \ref{tab:datasets}) to assess their suitability and limitations in the context of modern malware detection. Despite their popularity, these datasets present several important limitations when compared with the MH-1M dataset.

First, they are generally limited in scale. The largest among them, KronoDroid, contains around 78,000 samples, while MH-1M includes more than 1.34 million, making it an order of magnitude larger. This broader scale enhances the capacity of deep learning models to generalize across diverse real-world applications.

Second, most existing datasets have a narrow feature scope. Many, such as ADROIT and Android Permissions, focus on a single type of feature, while others, like DREBIN and DefenseDroid, offer a broader mix. However, none match the comprehensive feature coverage of MH-1M, which includes SHA256 hashes, VirusTotal scan results, 407 intents, 214 permissions, 232 opcodes, and over 22,000 API call features. This diversity enables more detailed and fine-grained analyzes of application behavior.

Third, several datasets rely on limited or outdated malware taxonomies, often lacking transparent labeling criteria. In contrast, MH-1M benefits from detailed VirusTotal-based labeling, allowing customizable thresholds and confidence levels that improve label reliability and experimental reproducibility.

Another critical limitation is the imbalance and noise found in existing datasets. For instance, Android Permissions show a skewed distribution (20,000 malware versus 9,999 benign samples), while Adroit exhibits significant unevenness in feature representation. Smaller datasets also tend to suffer from data sparsity, which reduces their usefulness for large-scale model training and generalization across app categories. In contrast, MH-1M maintains a realistic malware-to-benign ratio of approximately 1:10, reflecting real-world distributions and preventing overfitting to synthetic or overly balanced datasets.

Finally, MH-1M provides clean, standardized metadata and consistent identifiers, such as package names, file names, and hashes, which are often missing or inconsistent in other datasets. This structure facilitates integration with external threat intelligence sources and enhances traceability and cross-referencing of samples.


Our contribution is threefold.
First, we introduce the structured MH-1M dataset, which comprises 1,340,515 Android applications and integrates a diverse range of key features for malware detection, including intents, permissions, opcodes, and API calls.
Second, we provide open-access GitHub, Figshare, and Harvard Dataverse repositories containing more than 400 GB of supplementary material, featuring the most extensive collection of metadata ever assembled for a malware detection dataset.
Third, we present a preliminary analysis based on VirusTotal labeling, offering valuable insights into malware applications, their families, and behavioral groups.
In addition, we discuss and compare MH-1M with the MH-100K dataset to highlight the progress and improvements achieved in terms of scale, richness, and research applicability.

\section*{Methods}



The MH-1M dataset was developed using three primary tools: ADBuilder \cite{vilanova2022adbuilder, malwarehunter2023adbuilder}, AMGenerator \cite{rocha2023am_generator_explorer, malwarehunter2023amgenerator}, and AMExplorer \cite{rocha2023am_generator_explorer, malwarehunter2023amexplorer}.
ADBuilder represents an early effort to automate the creation of datasets for Android malware detection. Its architecture is composed of four main modules: i) downloading Android application packages (APKs), ii) extracting static features, iii) labeling samples as benign or malicious, and iv) generating the final dataset.
Although ADBuilder laid the foundation for automated dataset generation, it presented two major limitations that affected the reliability and completeness of the resulting datasets. The first limitation involved the labeling module, which relied exclusively on cached analysis results from the VirusTotal platform. This dependency often led to outdated or inaccurate classifications, as many applications had not been reassessed for several years. The second limitation concerned the restricted output format, which generated only binary datasets based on conventional feature categories such as permissions, API calls, intents, and opcodes, while omitting the broader range of metadata necessary for more detailed and reproducible analyzes.

\begin{figure*}[ht] 
    \begin{center} 
    \includegraphics[width=.7\textwidth]{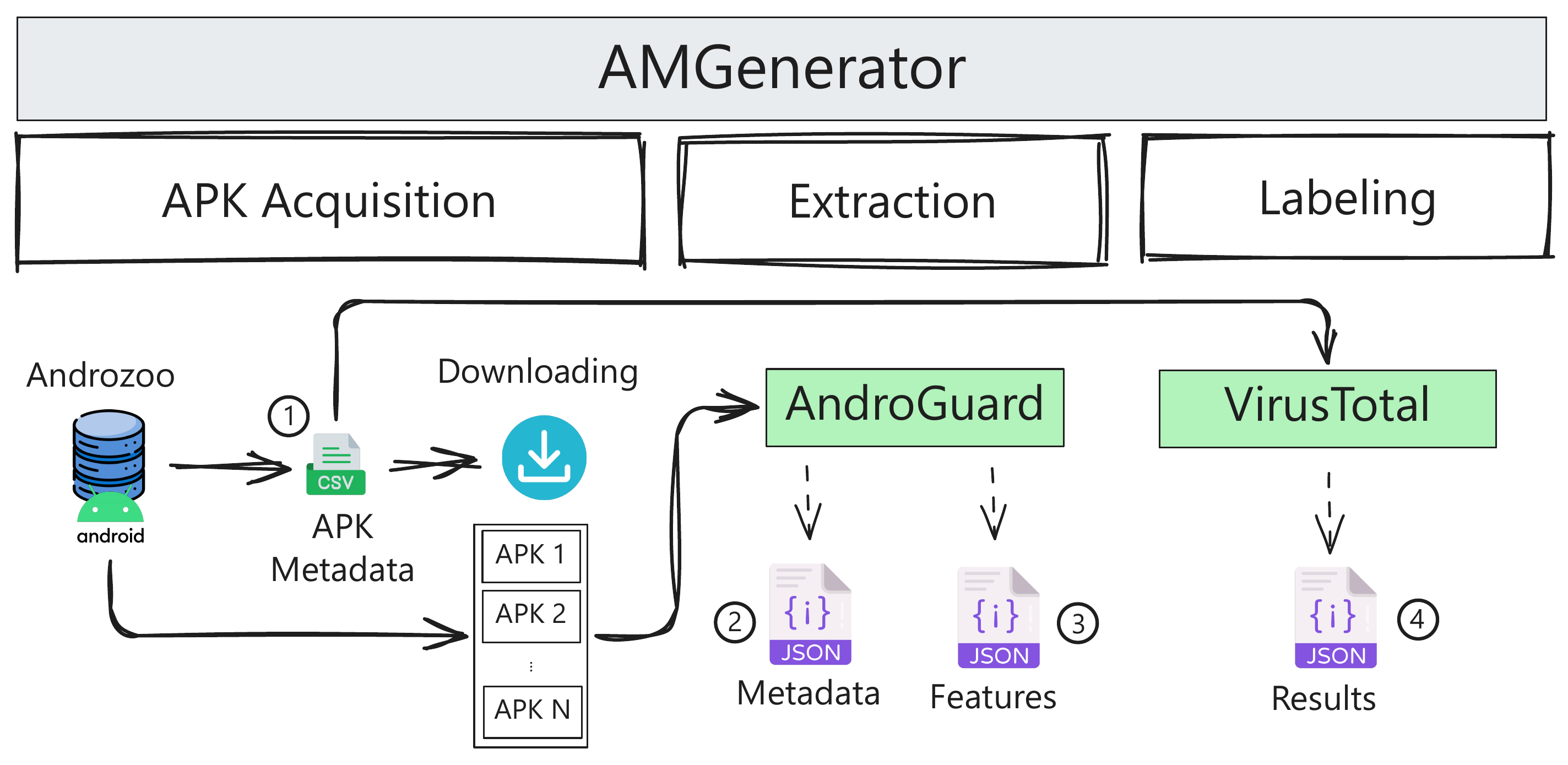} \end{center} 
    \caption{The overview of the AMGenerator tool, comprising acquisition, extraction, and labeling modules. The acquisition module retrieves APK files, the extraction module analyzes their features, and the labeling module assigns malware labels based on VirusTotal metadata.} 
    \label{fig:amgenerator} 
\end{figure*}

AMGenerator \cite{malwarehunter2023amgenerator} was developed as a direct evolution and refactoring of ADBuilder, specifically designed to overcome the limitations of its predecessor. The architecture of AMGenerator is illustrated in Figure \ref{fig:amgenerator}. It integrates the APK acquisition and feature extraction functionalities of ADBuilder, and its main advancement lies in a fully redesigned and more robust labeling module. This new module addresses the critical issue of data timeliness by ensuring that all labeling information remains current. The AMGenerator tool consists of three main modules: i) acquisition, ii) extraction, and iii) labeling.

The acquisition module receives as input a list of cryptographic hashes (for example, SHA256) of the target applications and attempts to download each APK from available sources. For the MH-1M dataset, only the AndroZoo repository \cite{allix2016androzoo} was used. All successfully downloaded applications are placed in the feature extraction queue. Any application whose download fails at this stage is discarded and excluded from subsequent dataset construction.

The extraction module processes the APKs from the acquisition queue and extracts characteristics from each one using specialized tools such as AndroGuard for static analysis. This phase is computationally demanding, as it involves unpacking the APK and generating intermediate code representations for analysis, including the construction of API call graphs. Once the extraction phase is complete, the resulting data file, containing the extracted features of each application, is passed to the data assembly pipeline for dataset construction.

The labeling module uses online services to obtain up-to-date information about each APK. In this work, we employ the VirusTotal (VT) platform to retrieve the labeling information for the MH-1M dataset. Online services such as VirusTotal aggregate the results of multiple antivirus scanners, and users must define a threshold (for example, five scanners) to classify an APK as benign or malicious. The VirusTotal service API \cite{virustotal2025api} is among the most comprehensive and widely used in this domain, as it integrates the results of more than 65 malware detection engines. The retrieved results are used to identify potentially malicious files and URLs.

Each VirusTotal API request returns a JSON file containing metadata specific to a single sample. This information can be used to classify the sample based on the number of scanners that flag the APK as potentially harmful. This approach enables a more accurate estimation of the threat level associated with each application, which is essential for a wide range of malware detection methods. Users may assign labels to samples using direct threshold-based counting or by adopting automated labeling solutions such as Maat \cite{salem2021maat}, which analyzes the metadata generated by VirusTotal to automatically determine the appropriate classification for each sample.

The current implementation of the labeling module accepts two input parameters: a list of SHA256 hashes that uniquely identify the APKs, and a valid VirusTotal API key. The final output for each analyzed APK is a JSON file that contains general metadata about the application, such as its name and extracted features, along with the results returned by all available VirusTotal scanners, indicating whether the sample is classified as benign or malicious.

\begin{figure*}[ht] 
    \begin{center} 
    \includegraphics[width=0.7\textwidth]{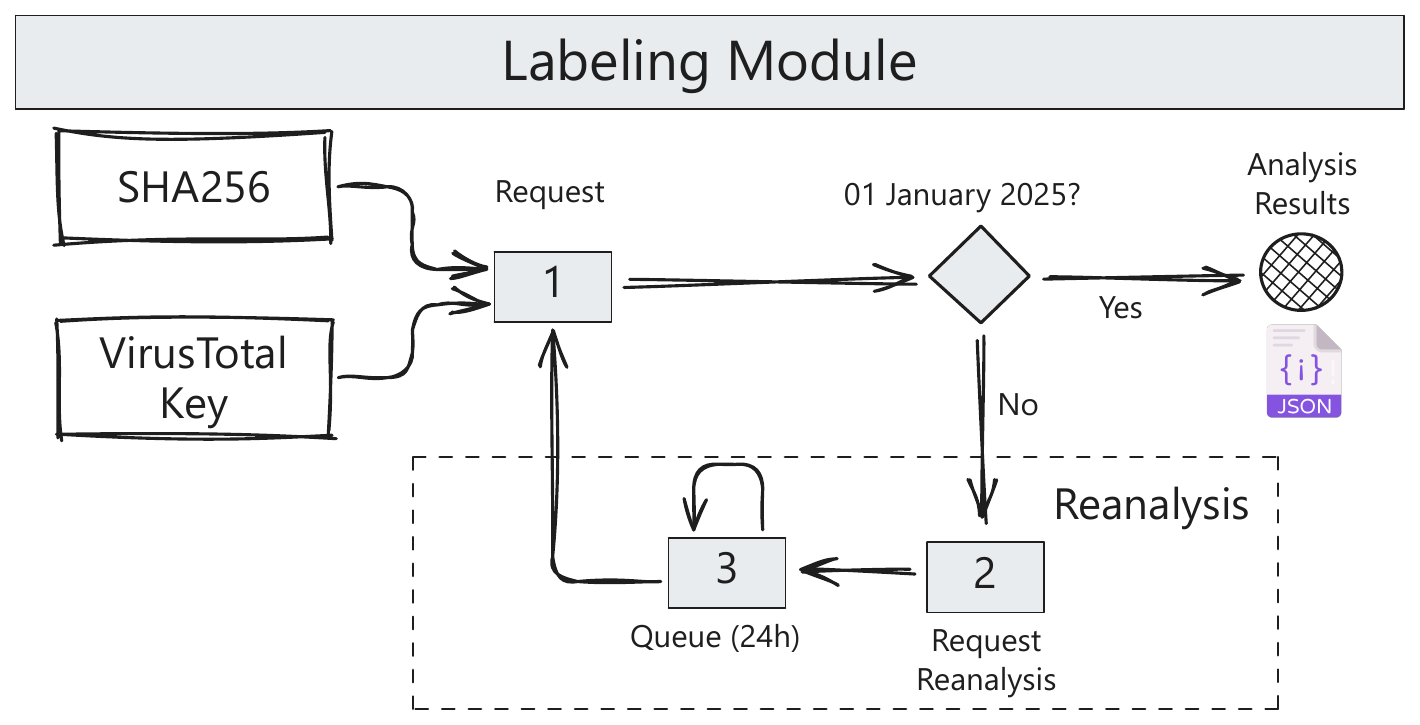} \end{center} 
    \caption{The process of obtaining and updating APK labels using VirusTotal metadata. If an APK was recently analyzed, metadata is retrieved directly; otherwise, a reanalysis request is submitted, ensuring up-to-date classification.} 
\label{fig_rotulacao} \end{figure*}

Figure \ref{fig_rotulacao} illustrates the metadata acquisition process for labeling performed through the VirusTotal platform. For APKs that have been recently analyzed by VirusTotal (for example, after a user-defined date such as January 1, 2025), a single request is sufficient to retrieve and use the metadata from the most recent analysis available in the cache. The result of this request is then stored in the labeling metadata repository.

If an application has never been analyzed or was analyzed prior to the user-defined cut-off date (for example, December 12, 2021), three separate requests are required: (1) a request for existing data, (2) a reanalysis request, and (3) a follow-up request after a predefined waiting period, typically 24 hours. This process is depicted in Figure \ref{fig_rotulacao}. In practice, most APKs require reanalysis, as observed in our labeling data collection, which involves thousands of samples. As a result, the maximum number of APK analyzes that can be performed per day with a single VirusTotal API key is limited to fewer than 200. To scale this process, additional API keys or specific contractual agreements with VirusTotal are required. This procedure is essential to ensure that the metadata used for labeling remain accurate and up to date.

\begin{figure*}[h] 
    \begin{center} 
    \includegraphics[trim={0 3cm 0 0}, width=.7\textwidth]{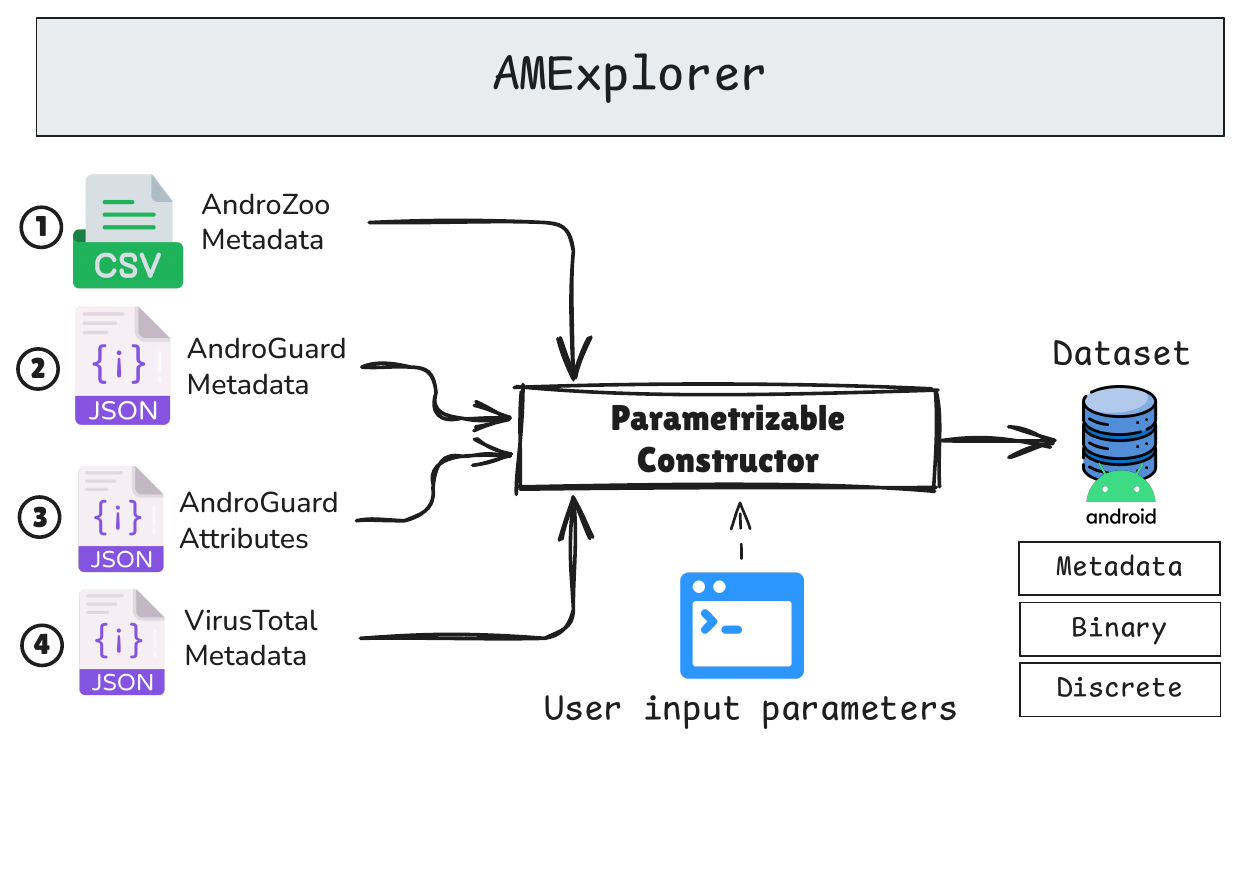} \end{center} 
    \caption{The AMExplorer tool aggregates and organizes metadata and feature sets extracted by AMGenerator. It generates three datasets: metadata, discrete, and binary, facilitating malware analysis and classification.} 
    \label{fig:amexplorer} 
\end{figure*}

The AMExplorer tool \cite{malwarehunter2023amexplorer, rocha2023am_generator_explorer} was developed to operate in conjunction with AMGenerator, focusing on the exploration and compilation of collected data into comprehensive and customizable datasets. AMExplorer provides users with the flexibility to generate different dataset types according to specific research objectives. As illustrated in Figure \ref{fig:amexplorer}, AMExplorer processes and integrates the outputs from the three AMGenerator modules (Figure \ref{fig:amgenerator}): (a) the acquisition module, which provides metadata obtained from the AndroZoo repository (1); (b) the extraction module, which supplies metadata (2) and attribute data (3) extracted using AndroGuard; and (c) the labeling module, which contributes the resulting VirusTotal metadata (4).

The AMExplorer tool (Figure \ref{fig:amexplorer}) includes a configurable set of input parameters that allows users to produce three distinct outputs. The first output is a metadata dataset, which consolidates the full collection of metadata for general analysis and tabulation purposes. The second and third outputs are discrete and binary datasets that are specifically structured for use in standard machine learning workflows. To generate these datasets, users must define a classification threshold, such as the minimum number of positive scanner detections required to label an application as malicious. Through this process, AMExplorer transforms the raw and continuously updated data produced by AMGenerator into organized and ready-to-use datasets suitable for advanced studies in Android malware detection.

\subsection*{MH-1M Use Cases }

The MH-1M dataset enables the investigation of a broad spectrum of contemporary research questions in the field of Android malware detection. Table \ref{tab:usecases_mh1m} summarizes key research trends and illustrates how MH-1M supports progress in each area.

\begin{table*}[p]
\centering
\caption{Research focuses and their alignment with the MH‑1M dataset.}
\label{tab:usecases_mh1m}
\begin{tabular}{@{}p{5cm} p{9cm}@{}}
\toprule
\textbf{Research Focus} & \textbf{Contribution to MH‑1M Use Case} \\
\midrule
Longitudinal Trends \cite{Janovsky2022Longitudinal, Daoudi2021Lessons} & Enables temporal analysis of API call and opcode usage changes over time, as MH‑1M includes metadata spanning more than a decade. \\
\addlinespace
Temporal Generalization \cite{Mariconti2016Mamadroid, Zheng2025Learning, Fan2021Heterogeneous, AlSobeh2024Android} & Facilitates evaluating how models trained on older APKs generalize to newer threats, mitigating temporal and spatial bias. \\
\addlinespace
Feature Selection \& Dimensionality Reduction \cite{ Daoudi2021Lessons, MehrabiKoushki2022On, Sedano2016Characterization, Alam2024Revisiting, Moran2024Machine, Kouliaridis2020Improving} & Supports investigation into which subsets of the 23,247 static features are most predictive, enabling deep exploration of reduction techniques such as PCA, LDA in large-scale contexts. \\
\addlinespace
Imbalance Handling \& VT Threshold Sensitivity \cite{Daoudi2021Lessons, Lin2022Dataset} & Supports rigorous examination of performance across different VirusTotal threshold settings and imbalance handling. \\
\addlinespace
Debiasing \& Representative Dataset Construction \cite{Lin2022Dataset} & Assists evaluation of debiasing strategies and subsampling techniques, leveraging MH‑1M’s large scale and balanced composition, and addressing dataset bias issues. \\
\addlinespace
Labeling Robustness \cite{miranda2022debiasing, Lin2022Dataset, Wang2022Malwhiteout} & Provides comprehensive VirusTotal metadata enabling exploration of both threshold-based and ML-based labeling strategies, thus improving labeling accuracy over time. \\
\bottomrule
\end{tabular}
\end{table*}

Because the dataset spans more than a decade and includes metadata such as compilation dates, it supports longitudinal studies that examine the evolution of malware behavior over time. This allows researchers to analyze the emergence of new malware families and shifts in feature prevalence, including changes in the distribution of API calls, opcodes, and permissions. With tens of thousands of static features, MH-1M also enables studies on the relative importance of different feature categories and the trade-offs involved in reducing feature space dimensionality.

MH-1M provides a comprehensive testbed for evaluating machine learning models under realistic deployment conditions. Its natural class imbalance, where malware represents approximately 9\% of all samples, facilitates the design and benchmarking of classification methods capable of effectively handling imbalanced data. The availability of up-to-date VirusTotal metadata also enables threshold sensitivity analyzes, allowing researchers to investigate how variations in detection thresholds (for example, one versus more than four detections) influence labeling quality and downstream model performance.

By structuring experiments with temporally stratified data splits, researchers can assess how models trained on earlier samples generalize to newer threats—an essential question for understanding the dynamics of the evolving malware ecosystem. This characteristic makes MH-1M particularly suitable as a benchmark for studying temporal concept drift and for evaluating the performance of continual learning strategies.

Beyond binary classification tasks, MH-1M supports more fine-grained research objectives, such as malware family classification and subgroup detection. When integrated with external malware taxonomies, the dataset allows for the training and evaluation of models capable of distinguishing between behaviorally similar but functionally distinct malware families. Its high dimensionality also facilitates the identification of detailed behavioral signatures embedded in API usage patterns and opcode sequences.

Finally, MH-1M serves as a valuable reference for research on dataset bias and representativeness. Its scale and diversity enable the validation of subsampling techniques and the development of bias mitigation strategies. It also provides a solid foundation for studies on data quality control, including analyzes of how duplicate or near-duplicate APKs affect feature distributions and model generalization.

\subsection*{Keeping MH-1M Up-to-date}

To keep the MH-1M dataset up-to-date and reflective of the evolving Android malware landscape, a versioning strategy grounded in temporal sampling was adopted. The dataset released is constructed by randomly selecting APKs samples from the AndroZoo repository, with selection constrained to a specific date range; for example, a version may include only APKs submitted during the third quarter of 2024 (Q3 2024). 
This time-based partitioning facilitates the organization and documentation of dataset versions (e.g., MH-1M\_2024Q4), and also enables reproducibility and supports longitudinal analyzes of malware evolution. 

However, this data selection approach introduces several potential sources of bias. First, temporal bias may arise due to the exclusive focus on specific collection periods, potentially leading to the omission of rare or emerging malware families that are not active within the selected timeframe. Second, source bias is inherent to AndroZoo itself, which primarily aggregates applications from Google Play and a limited set of alternative marketplaces. As a result, malware distributed through region-specific or unofficial channels may be underrepresented (e.g., third-party app stores, sideloading methods, or targeted campaigns). Third, reliance on VirusTotal for labeling introduces detection bias, as labels are derived from the consensus of multiple antivirus engines that may exhibit inconsistencies, false positives, or limited detection capabilities for novel threats. 

While random sampling is effective in mitigating systematic selection bias, it does not guaranty representation across all malware families or behavioral patterns unless stratification techniques are applied. To mitigate these limitations, future versions of MH-1M may benefit from enhancements such as stratified sampling based on malware families or behaviors, broader temporal coverage through overlapping windows, integration of APKs from a wider variety of sources, and, where feasible, manual curation or label refinement. Nevertheless, the current versioning strategy provides a scalable and transparent foundation for dataset maintenance, supporting robust and reproducible research in Android malware detection.

An important consideration when creating newer versions is the use of the VirusTotal (VT) service for labeling and metadata extraction. VirusTotal periodically updates its data schema to reflect changes in its internal analysis engines, added capabilities, or modifications in third-party integrations. As a result, key names, data formats, or nested structures in the VT reports may change over time, potentially affecting the consistency of parsed metadata or the stability of automated pipelines.

\section*{Data Records}

The MH-1M dataset is publicly available and can be accessed through the Figshare \cite{braganca2025} and GitHub \cite{hendrio2024mh1mGitHub} repositories, which host the processed data used in the experiments reported in this paper. The raw version of the MH-1M dataset, along with its extensive supplementary metadata exceeding 400 GB in size, is available at the Harvard Dataverse repository \cite{Braganca2025MH1M}. The dataset contains detailed information on 1,340,515 Android applications collected over a 14-year period (2010–2024). Among these, 91.1\% (1,221,421 samples) are benign, while 8.9\% (119,094 samples) correspond to various types of malware, providing a unique longitudinal perspective on malware evolution. The dataset includes a wide range of static features, comprising 407 intents, 214 permissions, 232 opcodes, and 22,394 API-call attributes.

The combination of API calls, intents, permissions, and opcodes enables a multilayered analysis of Android malware by integrating high-level behavioral indicators, such as intents and permissions, with low-level code characteristics captured through opcodes and API call sequences. Examples of permissions, intents, and API call features are presented in Table \ref{tab:features}.

\begin{table*}[h]
\centering
\caption{Examples of permissions, intents and api calls features used in MH-1M.}
\renewcommand{\arraystretch}{1.2}
\begin{tabular}{@{}l@{}}
\toprule
\multicolumn{1}{c}{\textbf{Features}}                                                 \\ \midrule
Permissions::WAKE\_LOCK                                                               \\
Permissions::WRITE\_EXTERNAL\_STORAGE                                                 \\
Permissions::ACCESS\_NETWORK\_STATE                                                   \\
Permissions::WRITE\_SETTINGS                                                          \\
Permissions::INTERNET                                                                 \\
Permissions::ACCESS\_WIFI\_STATE                                                      \\
Permissions::ACCESS\_FINE\_LOCATION                                                   \\
Permissions::ACCESS\_COARSE\_LOCATION                                                 \\
Permissions::READ\_PHONE\_STATE                                                       \\
Intents::AUDIO\_BECOMING\_NOISY                                                       \\
Intents::PACKAGE\_REPLACED                                                            \\
Intents::CONNECTIVITY\_CHANGE                                                         \\
Intents::WIFI\_STATE\_CHANGED                                                         \\
Intents::LOCKED\_BOOT\_COMPLETED                                                      \\
Intents::MY\_PACKAGE\_REPLACED                                                        \\
Intents::ACTION\_POWER\_CONNECTED                                                     \\
Intents::ACTION\_POWER\_DISCONNECTED                                                  \\
Intents::PHONE\_STATE                                                                 \\
APICalls::Landroid/content/Intent.toUri()                                             \\
APICalls::Landroid/view/View.setTag()                                                 \\
APICalls::Landroid/util/Xml.newSerializer()                                           \\
APICalls::Landroid/content/pm/PackageManager.queryIntentServices()                    \\
APICalls::Landroid/view/ViewStub.setLayoutResource()                                  \\
APICalls::Landroid/content/Context.createPackageContext()                             \\
APICalls::Landroid/os/Handler.removeMessages()                                        \\
APICalls::Landroid/widget/Scroller.getCurrY()                                         \\
APICalls::Landroid/view/View.invalidate()                                             \\
APICalls::Landroid/os/IBinder.linkToDeath()                                           \\
APICalls::Landroid/view/View\$AccessibilityDelegate.sendAccessibilityEventUnchecked() \\ \bottomrule
\end{tabular}
\label{tab:features}
\end{table*}

\begin{itemize}

\item \textbf{Intents} are a fundamental mechanism for inter-process communication in Android. They are used to start activities, services, or broadcast receivers, enabling an application to request actions from other components. Intents may be explicit, targeting a specific component, or implicit, declaring a general action that can be handled by any compatible component. In the context of malware analysis, examining intents can reveal attempts to invoke system services, initiate background processes, or trigger unauthorized activities.  

\item \textbf{Opcodes} correspond to the instructions used in Dalvik or ART (Android Runtime) bytecode, which forms the execution layer of Java-based Android applications. The analysis of these low-level instructions provides insights into an application's control flow and operational behavior. Opcode patterns, such as loops, conditional branches, and method invocations, can indicate obfuscation techniques or other behaviors associated with malicious activities.  

\item Android applications declare \textbf{permissions} in their manifest files to request access to system resources and sensitive user data. These permissions govern functionalities such as access to contacts, location, SMS messages, and device storage. In malware analysis, the set of requested permissions can serve as a strong indicator of potential misuse, particularly when the permissions requested appear excessive or unrelated to the stated functionality of the application.  

\item \textbf{API calls} represent the interactions between an application and the Android framework or third-party libraries. These calls define how the app communicates with device hardware, accesses system services, and performs network operations. By analyzing API call patterns, it is possible to infer application behavior, detect anomalous or suspicious activities, and identify potential attempts to exploit system vulnerabilities.  

\end{itemize}

The MH-1M dataset is stored in a compressed file ($.npz$) that contains several dictionary objects with detailed information about each application, as summarized in Table \ref{tab:file_keys}. The dataset includes metadata for each application instance, such as the SHA256 hash (APK signature), file name, package name, and an extensive collection of files containing the processed outputs from the VirusTotal analysis, as presented in Table \ref{tab:metadata}. To access the complete set of outputs, refer to the MH-1M raw data section. The additional information provided is valuable for future studies that aim to analyze antivirus scan results to better understand the prevalence, distribution, and behavioral characteristics of various malware families.

\begin{table*}[h]
\centering
\caption{The MH-1M dataset is stored in a compressed archive ($.npz$) that contains the data, features and metadata.}
\renewcommand{\arraystretch}{1.2}
\begin{tabular}{@{}ll@{}}
\toprule
\textbf{Key}             & \textbf{Description}                                                                                                                                     \\ \midrule
data                     & \begin{tabular}[c]{@{}l@{}}Primary data, which represents the processed binary features extracted from malware samples.\end{tabular}  \\
sha256                   & \begin{tabular}[c]{@{}l@{}} Unique identifier SHA-256 hashes, used for file integrity and verification.\end{tabular} \\
feature\_names           & \begin{tabular}[c]{@{}l@{}}Lists the features names for the metadata dataframe.\end{tabular}                                                          \\
metadata                 & \begin{tabular}[c]{@{}l@{}}Includes additional information  about the malware samples.\end{tabular}                                                    \\
metadata\_feature\_names & \begin{tabular}[c]{@{}l@{}}Provides the descriptions for each column in the metadata dataframe.\end{tabular}                                          \\ \bottomrule
\end{tabular}
\label{tab:file_keys}
\end{table*}
\begin{table*}[h]
\centering
\small
\caption{Description of attributes of metadata array.}
\begin{tabular}{@{}ll@{}}
\toprule
\textbf{Name}                          & \textbf{Description}                                                                                                                                                                               \\ \midrule
SHA256                                 & \begin{tabular}[c]{@{}l@{}}The SHA-256 hash of the file used to verify integrity.   \\ This serves as a unique identifier for each sample.\end{tabular}                                            \\
VT\_LAST\_ANALYSIS\_DATE               & \begin{tabular}[c]{@{}l@{}}The date when VirusTotal last performed an analysis on the file. \\ This timestamp helps track the recency of  the analysis.\end{tabular}                               \\
VT\_SIZE                               & \begin{tabular}[c]{@{}l@{}}The size of the file (bytes) as reported by VirusTotal. \\ This information can be useful for filtering files based on their size.\end{tabular}                         \\
VT\_MD5                                & The MD5 hash of the file as provided by VirusTotal.                                                                                                                                                \\
VT\_TIMES\_SUBMITTED                   & \begin{tabular}[c]{@{}l@{}}The number of times the file has been submitted to VirusTotal. \\ A higher submission count might indicate a file of interest or concern.\end{tabular}                  \\
VT\_SCANNERS\_FAILURE                  & The count of VirusTotal scanners that failed to analyze the file.                                                                                                                                  \\
VT\_SCANNERS\_MALICIOUS                & \begin{tabular}[c]{@{}l@{}}The number of scanners that flagged the file as malicious. \\ This is a key indicator of the file’s threat level.\end{tabular}                                          \\
VT\_SCANNERS\_UNDETECTED               & \begin{tabular}[c]{@{}l@{}}The count of scanners that did not detect the file as malicious. \\ This can be compared with the malicious count to gauge consensus among scanners.\end{tabular}       \\
VT\_SCANNERS\_SUGGESTED\_THREAT\_LABEL & \begin{tabular}[c]{@{}l@{}}A suggested threat label based on the analysis by VirusTotal scanners. \\ This label may indicate the type of  malware or threat associated with the file.\end{tabular} \\
VT\_SCANNERS\_NAMES                    & \begin{tabular}[c]{@{}l@{}}The names of the VirusTotal scanners that analyzed the file. \\ This provides context on which antivirus engines contributed to the analysis.\end{tabular}              \\
YEAR                                   & \begin{tabular}[c]{@{}l@{}}The year when the file was analyzed or when it was added to the dataset. \\ This can help in temporal analysis of threats.\end{tabular}                                 \\
CLASS                                  & The classification label that denote whether the file is benign  (0) or malicious (1)                                                                                                              \\ \bottomrule
\end{tabular}
\label{tab:metadata}
\end{table*}

The function presented in Algorithm \ref{alg:mh_load} provides a method to load the MH-1M dataset. It begins by loading an $.npz$ file using \textit{NumPy}'s $np.load$ method. Once the file is loaded into the variable $data\_dict$, the function extracts the metadata array and converts it to a \textit{DataFrame}. This \textit{DataFrame} is created using the column names specified in the $``metadata\_feature\_names"$ key of the NPZ file, which provides contextual information about each sample.
Next, the function retrieves the names of the features (stored under the $``feature\_names"$ key) and the SHA-256 hash values (stored under the $``sha256"$ key) from the loaded dictionary. Finally, the function returns a tuple consisting of the main data array, the metadata (\textit{DataFrame}), the feature names, and the SHA-256 hashes. 
When the $data\_dict[``data"]$ variable is loaded, it can use up to 30 GB of memory. This high usage of memory occurs because the data array is large, both in terms of the number of samples and features. Consider using techniques such as memory mapping $(np.load(..., mmap\_mode=``r"))$ if you need to work with data that cannot fully reside in memory, or explore other data processing libraries (e.g., Dask or Vaex).


\begin{algorithm*}[h]
\caption{Load MH-1M Malware Dataset.}
\begin{algorithmic}[1]
\Procedure{load\_mh}{path, file\_name}
    \State data\_dict $\gets$ \texttt{np.load(join(path, file\_name), allow\_pickle=True)}
    \State metadata $\gets$ \texttt{dataframe(data\_dict["metadata"], \newline columns=data\_dict["metadata\_feature\_names"])}
    \State columns\_names $\gets$ \texttt{data\_dict["feature\_names"]}
    \State sha256 $\gets$ \texttt{data\_dict["sha256"]}
    \State \textbf{print} \texttt{data\_dict["data"].shape}
    \State \Return \texttt{data\_dict["data"]}, metadata, columns\_names, sha256
\EndProcedure
\end{algorithmic}
\label{alg:mh_load}
\end{algorithm*}

\subsection*{MH-1M Android Malware Families}

In VirusTotal reports, malware family names are typically standardized representations derived from antivirus (AV) engines (for example, \textit{Trojan.FakeApp.Dnotua}). These names often follow hierarchical naming conventions that describe the type, family, and variant of the malware, as illustrated in Figure \ref{tab:vt_families}. It is important to note that different AV engines adopt slightly different naming schemes, and VirusTotal aggregates these results. Some AV engines include additional segments, such as \textit{Trojan.AndroidOS.FakeApp.Dnotua}, to explicitly indicate the target platform (e.g., AndroidOS).

\begin{table*}[h]
\centering
\caption{Segments of a typical malware naming convention with their purposes and examples.}
\begin{tabular}{@{} l p{6cm} p{6cm} @{}}
\toprule
\textbf{Segment} & \textbf{Typical Purpose} & \textbf{Example} \\
\midrule
Platform & Specifies the operating system or environment targeted & AndroidOS, Win32, Linux, Java, SymbOS \\
Category/Type & Indicates the high-level classification of the threat & Trojan, Adware, Backdoor, Worm, Spyware, Dropper, Exploit, Riskware \\
Family Name & Defines the malware family or behavior group, usually based on functionality or shared codebase & FakeApp, Hiddad, Agent, Ewind, BankBot \\
Subtype/Variant & Identifies a specific strain or variation within a family, such as new obfuscations, payloads, or delivery methods & A, B, Dnotua, Gen, v1, dldr \\
Detection Method/Heuristics & Denotes how the malware was identified—often used for generic or heuristic detections & Gen, Heur, Generic, Obfuscated, Packed, Suspicious \\
Functional Tag & Additional behavioral or campaign tag (e.g., for SMS-sending, spyware module, ransomware, etc.) & SMSAgent, ClipBanker, SpyModule, Ransom, CoinMiner \\
\bottomrule
\end{tabular}

\label{tab:vt_families}
\end{table*}

The MH-1M dataset provides a robust foundation for the study and classification of malware families. By processing the associated metadata, malware samples have been grouped into broad superclasses of families, including adware, trojan, riskware, and other malware categories based on their behavioral characteristics and threat profiles. This categorization highlights the prevalence and diversity of different malware types, enabling longitudinal analyzes of their evolution and emerging trends over time. The dataset includes four principal categories of malware families, excluding benign samples, as summarized in Table \ref{tab:malware}.


\begin{table*}[h]
\centering
\small
\caption{The malware family distribution on MH-1M dataset.}
\begin{tabular}{@{}lccl@{}}
\toprule
\textbf{Malware Family} & \textbf{ID} & \textbf{Count} & \textbf{Description}                                                                                                                                               \\ \midrule
Adware                  & 1           & 59,065         & \begin{tabular}[c]{@{}l@{}}Software that typically displays unsolicited advertisements, \\ often generating revenue through ad impressions.\end{tabular}           \\
Trojan                  & 2           & 44,257         & \begin{tabular}[c]{@{}l@{}}Malware disguised as legitimate software, \\ designed to deceive users and facilitate unauthorized access.\end{tabular}                 \\
Riskware                & 3           & 9,022          & \begin{tabular}[c]{@{}l@{}}Programs that possess legitimate functionalities \\ but may also pose security risks due to vulnerabilities or misuse.\end{tabular}     \\
Others                  & 4           & 2,774          & \begin{tabular}[c]{@{}l@{}}A catch-all category for malware that \\ does not fit into the above classifications or \\ exhibits mixed characteristics.\end{tabular} \\ \bottomrule
\end{tabular}
\label{tab:malware}
\end{table*}

\subsection*{MH-1M Raw Data}
\label{sec:rawdata}

To ensure the long-term preservation, accessibility, and discoverability of the research data, all raw datasets used in this study have been formally archived in the MH-1M Harvard Dataverse repository \cite{Braganca2025MH1M}. We encourage other researchers to access and use this data for replication and further investigation.

The MH-1M raw data repository is organized into three primary components: features, labels, and metadata. Due to the large size of the dataset, the feature and label archives have been compressed and divided into multiple smaller files to facilitate easier download and management. As shown in Table \ref{tab:dataverse_repository}, the feature data are divided into 23 parts, and the label data are split into 6 parts. The metadata are contained in a single, smaller archive. To reconstruct the complete archives for features and labels, users should download all corresponding parts and concatenate them before decompressing the resulting $.tar.gz$ file.

\begin{table*}[h]
\centering
\caption{Summary of MH-1M raw data repository files stored in Harvard Dataverse. 
}
\begin{tabular}{@{}llll@{}}
\toprule
\textbf{Name}          & \textbf{Start Index} & \textbf{End Index} & \textbf{Total Size (GB)} \\ \midrule
features.tar.gz.part\_ & 001                  & 023                & 45.6                     \\
label.tar.gz.part\_    & 001                  & 006                & 11.5                     \\
metadata.tar.gz        & \multicolumn{2}{c}{N/A}                   & 0.13                     \\ \bottomrule
\end{tabular}%
\label{tab:dataverse_repository}
\end{table*}

The Table \ref{tab:vt_report} provides an overview of the core structural elements in a VirusTotal file report. 
The attributes field in a VirusTotal file report encapsulates rich metadata and scan results essential for malware analysis and labeling. 
It contains verified hash values (\textit{md5}, \textit{sha1}, \textit{sha256}), file characteristics (\textit{type\_description}, \textit{size}, \textit{meaningful\_name}), timestamps for submission and analysis (\textit{first\_submission\_date}, \textit{last\_submission\_date}, \textit{last\_analysis\_date}), community-based reputation scores, tags, and inferred features such as packers and sandbox verdicts.

\begin{table*}[h]
\centering
\caption{VirusTotal report information for an Android APK.}
\label{tab:vt_report}
\begin{tabular}{@{}p{3cm} p{11cm}@{}}
\toprule
\textbf{Key} & \textbf{Value Description} \\
\midrule
\texttt{data} & The top-level container holding the file object, including nested \texttt{type}, \texttt{id}, \texttt{attributes}, and optionally \texttt{relationships}. Contains all relevant analysis results. \\
\addlinespace
\texttt{type} & A string indicating the object’s type, commonly `"file"` for APKs, reflecting the API v3 object schema.  \\
\addlinespace
\texttt{id} & A unique identifier for the object, which in the case of a file object corresponds to the SHA256 hash of the APK.  \\
\addlinespace
\texttt{attributes} & A nested dictionary that contains detailed metadata and analysis—such as hashes, scan results, reputation, tags, and file statistics. This is the primary payload for researchers.  \\
\bottomrule
\end{tabular}
\end{table*} 

Table \ref{tab:vt_attributes} presents variables under  \textit{attributes}. The \textit{last\_analysis\_stats} variable provides aggregate counts of antivirus engine outcomes (malicious, harmless, suspicious, undetected, etc.), while the \textit{last\_analysis\_results} variable offers granular per-engine verdicts, including keys such as \textit{category}, \textit{engine\_name}, \textit{method}, \textit{result}, \textit{engine\_version}, and \textit{engine\_update}. An example of the Kaspersky engine result can be seen in Table \ref{tab:vt_kaspersky_analysis}.


\begin{table*}[h]
\centering
\caption{Important variables under \texttt{attributes} in VirusTotal report.}
\label{tab:vt_attributes}
\renewcommand{\arraystretch}{1.2}
\begin{tabular}{@{}lll@{}}
\toprule
\textbf{Key} & \textbf{Value Type} & \textbf{Description} \\ \midrule
\texttt{last\_analysis\_stats} & \texttt{dict} & Count of engines: malicious, harmless, suspicious, etc. \\
\texttt{last\_analysis\_results} & \texttt{dict} & Detailed results from each antivirus engine \\
\texttt{md5}, \texttt{sha1}, \texttt{sha256} & \texttt{str} & Cryptographic hashes of the APK file \\
\texttt{type\_description} & \texttt{str} & File type description, e.g., APK or Android app \\
\texttt{meaningful\_name} & \texttt{str} & File name or inferred name \\
\texttt{size} & \texttt{int} & File size in bytes \\
\texttt{first\_submission\_date} & \texttt{int} & Timestamp of first submission to VirusTotal \\
\texttt{last\_submission\_date} & \texttt{int} & Timestamp of last analysis \\
\texttt{reputation} & \texttt{int} & Community-based reputation score \\
\texttt{tags} & \texttt{list[str]} & Tags like ``android'', ``trojan'', etc. \\
\texttt{popular\_threat\_classification} & \texttt{dict} & Consensus threat label and common name \\
\texttt{sandbox\_verdicts} & \texttt{dict} & Dynamic behavior verdicts from sandboxes \\
\texttt{exiftool} & \texttt{dict} & Extracted metadata (e.g., version code, SDK info) \\
\texttt{packers} & \texttt{list[str]} & Detected obfuscation/packing tools \\
\texttt{trid} & \texttt{list[dict]} & File type signatures using TRiD tool \\
\texttt{times\_submitted} & \texttt{int} & Submission count of this APK hash \\
\bottomrule
\end{tabular}
\end{table*}

\begin{table*}[h]
\centering
\caption{Example of \texttt{last\_analysis\_results} Entry for Kaspersky}
\renewcommand{\arraystretch}{1.2}
\begin{tabular}{@{}ll@{}}
\toprule
\textbf{Key} & \textbf{Value} \\ \midrule
\texttt{category} & malicious \\
\texttt{engine\_name} & Kaspersky \\
\texttt{method} & blacklist \\
\texttt{result} & HEUR:Trojan.AndroidOS.Boogr.gsh \\
\texttt{engine\_version} & 21.3.62.0 \\
\texttt{engine\_update} & 2025-07-31 \\
\bottomrule
\end{tabular}
\label{tab:vt_kaspersky_analysis}
\end{table*}

This dual-structure approach supports robust labeling strategies: the aggregate \textit{last\_analysis\_stats} enables threshold-based labeling (e.g., requiring $> 4$ engines to be flagged as malicious), whereas the detailed \textit{last\_analysis\_results} allows for accounting for vendor-specific behaviors or for weighing engines differently in ensemble classification. 

\section*{Technical Validation}

We conduct two main evaluations using the MH-1M dataset. The first is an assessment of the MH-1M dataset itself. The second is a comparative study that builds upon our previously developed MH-100K dataset \cite{bragancca2023capturing}.

To construct an Android malware classification model, we employ the XGBoost (Extreme Gradient Boosting) classifier, which is well known for its strong performance in tabular data classification \cite{shwartz2022tabular}. Despite recent advances in deep learning methods for tabular data, previous research indicates that XGBoost consistently achieves superior results across diverse datasets.

We apply the holdout validation method with a 70–30 data split and evaluate model performance using standard classification metrics, including accuracy, precision, recall, F1-score, and the confusion matrix. In addition, following our previous findings \cite{bragancca2023capturing, braganca2023xai}, we adopt four scanners as the optimal threshold for defining malware labels in our classification task. The following section provides a detailed discussion of this labeling strategy.

\subsection*{Labelling Android Applications}

Before presenting the details of our labeling process, it is important to understand that the VirusTotal service does not provide a single definitive label, such as ``malware'' or ``benign'', for a submitted APK.

VirusTotal analyzes Android APK files using dozens of antivirus (AV) engines from various security vendors, including Kaspersky, Bitdefender, Avast, and others. Each engine independently examines the file and returns a verdict, either marking it as malicious with a specific malware label (for example, \textit{Trojan.Android.Agent} or \textit{Adware.Leadbolt}) or classifying it as clean or undetected. As a result, the outputs from these scanners may differ, leading to cases where an APK is flagged as malicious by some engines while being considered legitimate by others.

The number of scanners that detect an APK as malware represents how many of these engines have flagged the file as malicious. For instance, if 34 out of 72 engines report a positive detection, VirusTotal will indicate ``34/72 engines detected this file as malicious''. This detection count is commonly used as a proxy for label confidence in malware analysis. A file detected by only one or two engines may represent a false positive, whereas a file flagged by several engines (for example, eight or more) is far more likely to be truly malicious \cite{bragancca2023capturing}.

There is considerable variability among antivirus engines, with some adopting more aggressive detection strategies while others are more conservative. This variability requires researchers to apply a decision criterion by interpreting the set of results and defining a threshold, such as the minimum number of positive detections, which is used to classify each sample as either benign or malicious.

In our previous studies \cite{bragancca2023capturing}, we systematically varied the threshold for labeling an application as malware according to the number of antivirus scanners that flagged it. Specifically, we constructed eight different versions of the dataset, each corresponding to a distinct minimum number of scanners (from 1 to 8) required to label a sample as malicious. For example, in the version where the threshold is set to 8, an APK is labeled as malware only if at least eight scanners detect it as such.

After constructing these dataset variants, we trained an XGBoost classification model on each one to evaluate which threshold provided the best overall performance. We observed that increasing the number of required scanners improves label confidence but significantly reduces recall.

Based on our empirical results, we recommend using $detections \geq 4$ as an effective threshold for labeling malware. Accordingly, the variable $CLASS$ is assigned a value of $1$ (malware) when $VT_SCANNERS_MALICIOUS \geq 4$.

\subsection*{Android Malware Classification}

The quality of a dataset is largely an intrinsic property, independent of model architecture, class balance, or dataset size \cite{couch2025xfactor,molecules26041111}. However, classifier performance can serve as an indirect yet compelling validation of dataset quality, wherein the interpretation of results indicates whether the dataset is informative, clean, balanced, and suitable for predictive modeling.

When high-performing models achieve strong and consistent metrics such as precision, recall, F1-score, and AUC, this performance suggests that the features are discriminative, the labels are reliable, and the class separability is sufficient to support learning. Conversely, poor performance often reflects issues such as label noise, feature ambiguity, limited representativeness, or concept overlap, all of which indicate potential deficiencies in the dataset.

Extensive empirical evidence from Android malware detection research consistently shows that ensemble machine learning models such as Random Forest \cite{biau2016random}, XGBoost \cite{chen2015xgboost}, and CatBoost \cite{prokhorenkova2018catboost} often outperform deep learning architectures when applied to structured static or behavioral features \cite{muzaffar2024investigating, liu2025benchmarking}. More broadly, studies on tabular data demonstrate that classical machine learning models frequently surpass deep learning methods in performance \cite{grinsztajn2022tree, shwartz2022tabular, mcelfresh2023neural}. Deep learning approaches do not necessarily outperform traditional models, particularly in structured feature scenarios such as malware classification. Moreover, deep learning models typically require more extensive tuning and computational resources while still underperforming XGBoost in most tabular classification tasks.

Recent transformer-based models for tabular data, such as TabPFN \cite{hollmann2025accurate}, attempt to close this performance gap. However, their competitive advantage remains limited to smaller datasets, and they continue to lag behind tree-based models in large-scale, real-world applications \cite{mcelfresh2023neural, hollmann2025accurate}.

\begin{figure*}[h]
    \begin{minipage}[b]{0.5\textwidth}
        \centering
        \begin{subfigure}[b]{0.7\textwidth}
            \centering
            \includegraphics[width=\textwidth]{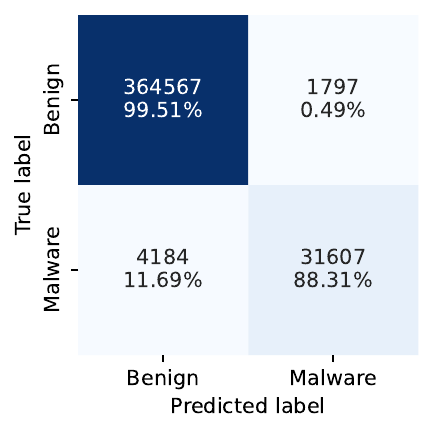}
            \caption{MH-1M}
            \label{fig_report_amex}
        \end{subfigure}
    \end{minipage}%
    \begin{minipage}[b]{0.5\textwidth}
        \centering
        \begin{subfigure}[b]{0.7\textwidth}
            \centering
            \includegraphics[width=\textwidth]{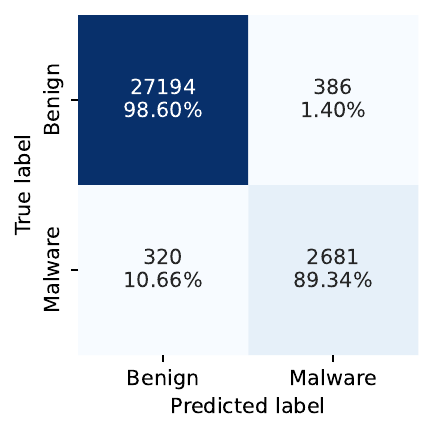}
            \caption{MH100K}
            \label{fig_report_mh100}
        \end{subfigure}
    \end{minipage}
    \caption{Confusion matrix results for XGBoot classifier.}
    \label{fig_comparison_amex_mh100}
\end{figure*}

In this study, we systematically employ gradient boosting ensemble classifiers, including XGBoost \cite{chen2015xgboost} and CatBoost \cite{prokhorenkova2018catboost}, to leverage their well-documented effectiveness in malware detection tasks \cite{10837854,fajar2024enhancingphishingdetectionfeature}. As shown in Table \ref{tab_xgboost_amex_report}, the performance of the XGBoost classifier on the MH-1M dataset demonstrates its robustness in distinguishing between benign and malicious applications. Figure \ref{fig_report_amex} further shows that the misclassification rate for benign applications (class 0) incorrectly identified as malicious (class 1) is only 0.49\%, indicating a highly accurate classification. Moreover, the model exhibits strong malware detection capability, with a relatively low 11.69\% misclassification rate for malware applications classified as benign.

\begin{table*}[h]
    \centering
    \begin{minipage}{.5\textwidth}
        \centering
        \resizebox{\textwidth}{!}{
        \begin{tabular}{lcccc}
            \toprule
            Class & Precision & Recall & F1-Score & Support \\
            \midrule
            0 & 0.9887 & 0.9951 & 0.9919 & 366364 \\
            1 & 0.9462 & 0.8831 & 0.9136 & 35791 \\
            \midrule
            Accuracy & \multicolumn{4}{c}{0.9851} \\
            \midrule
            Macro Avg & 0.9674 & 0.9391 & 0.9527 & 402155 \\
            Weighted Avg & 0.9849 & 0.9851 & 0.9849 & 402155 \\
            \bottomrule
        \end{tabular}
        }
         \caption{XGboost on the MH-1M.}
        \label{tab_xgboost_amex_report}
    \end{minipage}%
    \hfill
    \begin{minipage}{.5\textwidth}
        \centering
        \resizebox{\textwidth}{!}{
        \begin{tabular}{lcccc}
            \toprule
            Class & Precision & Recall & F1-Score & Support \\
            \midrule
            0 & 0.9884 & 0.9860 & 0.9872 & 27580 \\
            1 & 0.8741 & 0.8934 & 0.8837 & 3001 \\
            \midrule
            Accuracy & \multicolumn{4}{c}{0.9769} \\
            \midrule
            Macro Avg & 0.9313 & 0.9397 & 0.9354 & 30581 \\
            Weighted Avg & 0.9772 & 0.9769 & 0.9770 & 30581 \\
            \bottomrule
        \end{tabular}
        }
        \caption{XGboost on the MH100K.}
        \label{tab_xgboost_mh100_report}
    \end{minipage}
\end{table*}

In the results for the MH-100K dataset, presented in Table \ref{tab_xgboost_mh100_report}, the XGBoost classifier achieved an overall accuracy of 97.69\%. The precision for benign applications (class 0) reached 98.84\%, with a recall of 98.60\% and an F1-score of 98.72\%. These results demonstrate the model's strong ability to correctly identify benign samples while maintaining a very low false-positive rate.

For malicious applications (class 1), the model achieved a precision of 87.41\%, a recall of 89.34\%, and an F1-score of 88.37\%. Although the model performed well overall, these results indicate some challenges in detecting all malicious samples, leading to a slightly lower precision compared with the benign class.

When applied to the MH-1M dataset, the XGBoost model achieved an overall accuracy of 98.51\%, surpassing the performance obtained with the MH-100K dataset. The precision for benign samples (class 0) was 98.87\%, with a recall of 99.51\% and an F1-score of 99.19\%, confirming the model’s high effectiveness in identifying benign applications with minimal misclassification.

Regarding malicious samples (class 1), the model achieved a precision of 94.62\%, a recall of 88.31\%, and an F1-score of 91.36\%. Although these metrics indicate strong performance, the slightly lower recall suggests that some malicious applications were misclassified as benign. Given that the MH-1M dataset contains ten times more data than MH-100K and exhibits greater variability, this result may reflect the broader diversity of malware samples included in the dataset.

Overall, these results highlight the superior performance of the XGBoost model on the MH-1M dataset, demonstrating its robustness and reliability across a larger and more heterogeneous collection of applications. The increased number of samples and the inclusion of extensive metadata appear to play a key role in enhancing model performance. Moreover, the availability of more than 400 GB of metadata provides valuable opportunities for researchers to further develop and refine both supervised and unsupervised machine learning models.

\subsection*{Cross Classification and Evaluation}

For the cross classification experiments, we used only the shared features present in both the MH-1M and MH-100K datasets. This subset includes a total of 250 intents, 166 permissions, 0 opcodes, and 11,545 API calls. The analysis of cross-dataset testing results between MH-1M and MH-100K reveals notable differences in model performance when using XGBoost and CatBoost classifiers.

In the first scenario, shown in Table \ref{tab_cross_classification_report_amex}, the model was trained on the MH-1M dataset and tested on the MH-100K dataset. The results indicate strong generalization performance, with the CatBoost model achieving slightly better results, as presented in Table \ref{tab:catboost_mh1m}.

\begin{table*}[h]
    \centering
    \begin{minipage}{.48\textwidth}
        \centering
        \resizebox{\textwidth}{!}{
        \begin{tabular}{lcccc}
        \toprule
        Class & Precision & Recall & F1-Score & Support \\
        \midrule
        0 & 0.9929 & 0.9717 & 0.9822 & 92134 \\
        1 & 0.7783 & 0.9344 & 0.8492 & 9800 \\
        \midrule
        Accuracy & \multicolumn{4}{c}{0.9681} \\
        \midrule
        Macro Avg & 0.8856 & 0.9530 & 0.9157 & 101934 \\
        Weighted Avg & 0.9722 & 0.9681 & 0.9694 & 101934 \\
        \bottomrule
    \end{tabular}
        }
        \caption{XGBoost classification report. We use MH-1M to train and MH100K to test.}
        \label{tab_cross_classification_report_amex}
    \end{minipage}%
    \hfill
    \begin{minipage}{.48\textwidth}
        \centering
        \resizebox{\textwidth}{!}{
        \begin{tabular}{lcccc}
        \toprule
        Class        & Precision & Recall & F1-Score & Support \\ \midrule
        0            & 0.9314    & 0.9989 & 0.9640   & 1221421 \\
        1            & 0.9561    & 0.2450 & 0.3901   & 119094  \\ \midrule
        Accuracy     & \multicolumn{4}{c}{0.9319}              \\ \midrule
        Macro Avg    & 0.9437    & 0.6220 & 0.6770   & 1340515 \\
        Weighted Avg & 0.9336    & 0.9319 & 0.9130   & 1340515 \\ \midrule
        \end{tabular}%
        }
        
       \caption{XGBoost classification report. We use MH100K to train and MH-1M to test.}
        \label{tab_cross_classification_report_mh100}
    \end{minipage}
\end{table*}

\begin{figure*}[h]
    \begin{minipage}[b]{0.5\textwidth}
        \centering
        \begin{subfigure}[b]{0.7\textwidth}
            \centering
            \includegraphics[width=\textwidth]{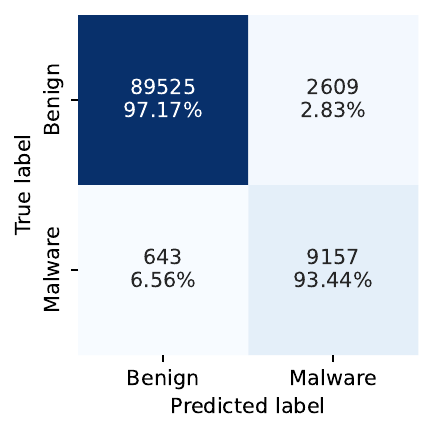}
            \caption{MH-1M}
            \label{fig:cross_amex}
        \end{subfigure}
    \end{minipage}%
    \begin{minipage}[b]{0.5\textwidth}
        \centering
        \begin{subfigure}[b]{0.7\textwidth}
            \centering
            \includegraphics[width=\textwidth]{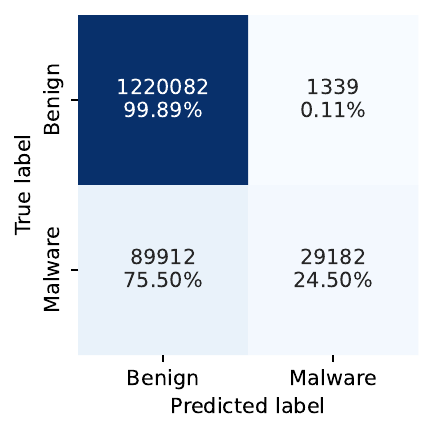}
            \caption{MH100K}
            \label{fig:cross_mh100}
        \end{subfigure}
    \end{minipage}
    \caption{Confusion matrix results for XGBoost classifier in cross classification.}
    \label{fig:cross_amex_mh100}
\end{figure*}

\begin{figure*}[h]
    \begin{minipage}[b]{0.5\textwidth}
        \centering
        \begin{subfigure}[b]{0.7\textwidth}
            \centering
            \includegraphics[width=\textwidth]{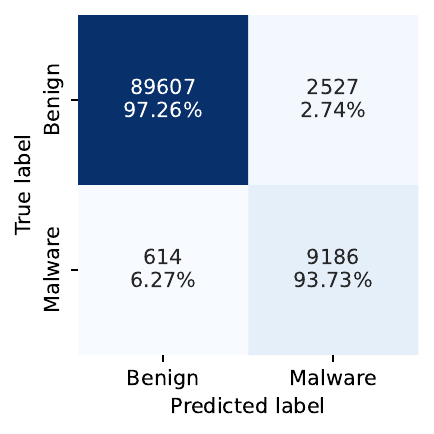}
            \caption{MH-1M}
            \label{fig:catboost_mh1m}
        \end{subfigure}
    \end{minipage}%
    \begin{minipage}[b]{0.5\textwidth}
        \centering
        \begin{subfigure}[b]{0.7\textwidth}
            \centering
            \includegraphics[width=\textwidth]{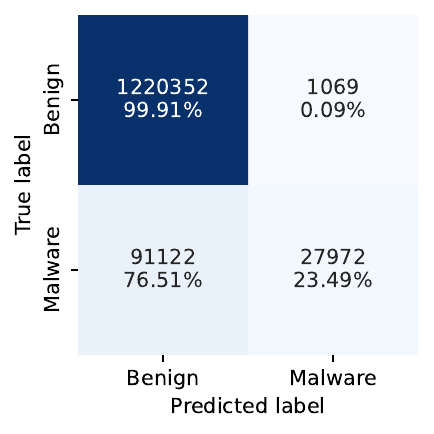}
            \caption{MH100K}
            \label{fig:catboost_mh100}
        \end{subfigure}
    \end{minipage}
    \caption{Confusion matrix results for CatBoost classifier in cross classification.}
    \label{fig:catboost_cv}
\end{figure*}

The outcomes illustrated in Figure \ref{fig:cross_amex} and Figure \ref{fig:catboost_mh1m} show a high degree of balance. Only 2.83\% of benign applications were misclassified as malware by the XGBoost model, and 2.74\% by the CatBoost model. Conversely, 6.56\% of malicious applications were misclassified as benign by the XGBoost model and 6.27\% by the CatBoost model.

In contrast, the second scenario (see Table \ref{tab_cross_classification_report_mh100} and Table \ref{tab:catboost_mh100}), where the models were trained on the MH-100K dataset and tested on the MH-1M dataset, yielded less favorable results. 
Specifically, the macro-recall average was only 62.20\%, compared to 95.30\% in the former case (see Table \ref{tab_cross_classification_report_amex}).
In fact, for the malicious class, the performance of the model was suboptimal, with 75.50\% of malware applications misclassified, as shown in Figure \ref{fig:cross_mh100}. 
This underscores the model's challenge in accurately classifying malicious samples from the larger and more diverse MH-1M dataset.

The performance disparity observed between the two cross-dataset experiments can be explained by several factors. Training the XGBoost model on the larger and more diverse MH-1M dataset provides a significant advantage by exposing the model to a broader range of benign and malicious samples. This diversity enhances the model’s ability to generalize and accurately classify samples from the MH-100K dataset. In contrast, when the model is trained on the smaller MH-100K dataset, it lacks exposure to the extensive variety of malware types and behaviors contained in MH-1M. As a result, its performance decreases when evaluated on the MH-1M dataset. These findings emphasize the critical importance of employing comprehensive and diverse datasets when training machine learning models for malware detection.


\begin{table*}[h]
\centering
\begin{minipage}{.5\textwidth}
\centering
\resizebox{\textwidth}{!}{
\begin{tabular}{lcccc}
\toprule
Class&Precision&Recall&F1-Score&Support\\
\midrule
0&0.9932&0.9726&0.9828&92134\\
1&0.7843&0.9373&0.8540&9800\\
\midrule
Accuracy&\multicolumn{4}{c}{0.9692}\\
\midrule
MacroAvg&0.8887&0.9550&0.9184&101934\\
WeightedAvg&0.9731&0.9692&0.9704&101934\\
\bottomrule
\end{tabular}
}
\caption{CatBoost classification report. We use MH-1M to train and MH100K to test.}
\label{tab:catboost_mh1m}
\end{minipage}%
\hfill
\begin{minipage}{.5\textwidth}
\centering
\resizebox{\textwidth}{!}{
\begin{tabular}{lcccc}
\toprule
Class&Precision&Recall&F1-Score&Support\\
\midrule
0&0.9305&0.9991&0.9636&1221421\\
1&0.9632&0.2349&0.3777&119094\\
\midrule
Accuracy&\multicolumn{4}{c}{0.9312}\\
\midrule
MacroAvg&0.9469&0.6170&0.6706&1340515\\
WeightedAvg&0.9334&0.9312&0.9115&1340515\\
\bottomrule
\end{tabular}
}
\caption{CatBoost classification report. We use MH100K (Train) and MH-1M (Test).}
\label{tab:catboost_mh100}
\end{minipage}
\end{table*}

\subsection*{Android Malware Families Analysis}

\begin{figure*}[h]
    \centering
    \includegraphics[ width=.9\textwidth]{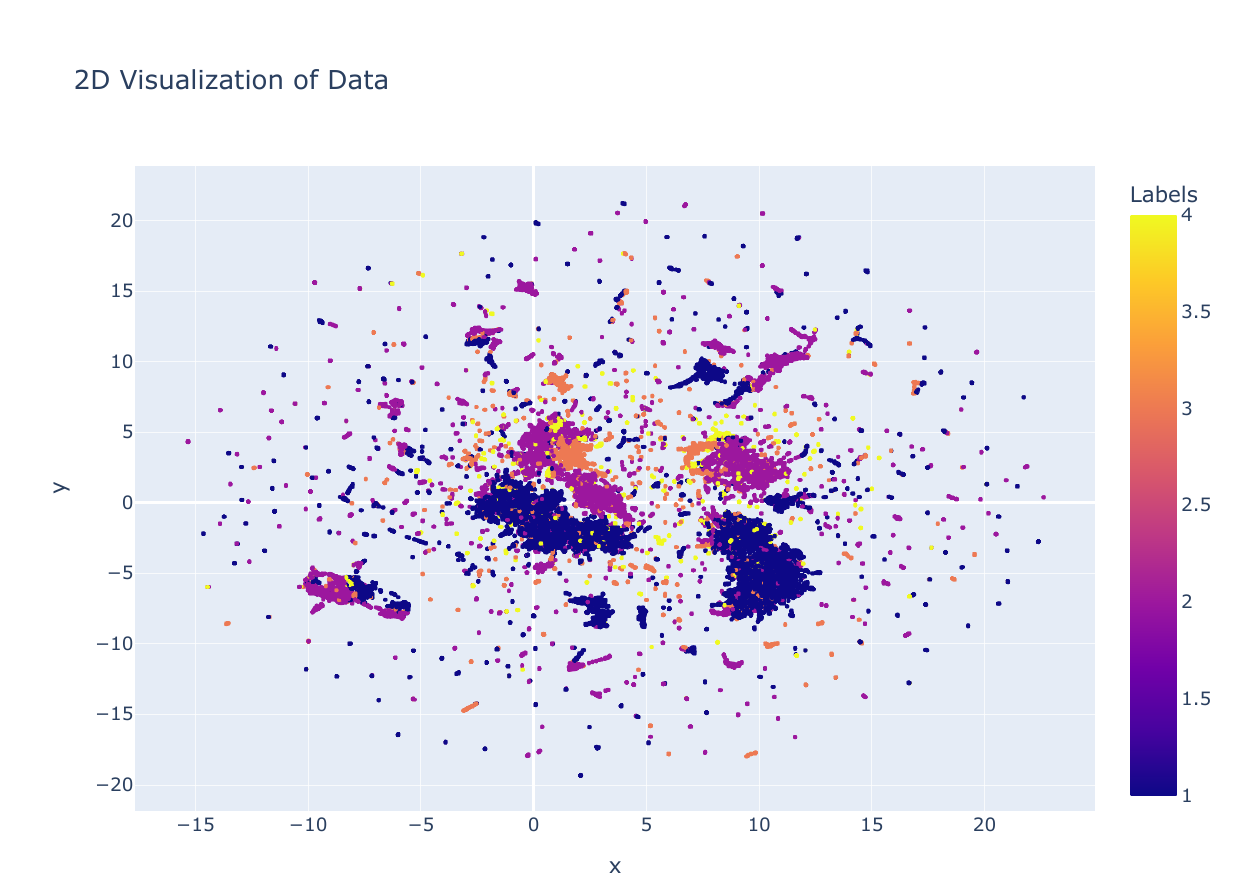}
    \caption{The UMAP projection maps malware samples into a two-dimensional latent space, where each point represents a malware APK and the color indicates its assigned class: Adware (1), Trojan (2), Riskware (3), or Others (4).}
    \label{fig:umap_families}
\end{figure*}

In an initial study, we presented a high-level classification of malware samples identified in the MH-1M dataset, as shown in Figure \ref{fig:umap_families}. The UMAP \cite{mcinnes2018umap} visualization, constructed from high-dimensional features such as API calls, permissions, intents, and opcodes, projects the malware samples into a two-dimensional latent space. In this representation, each point corresponds to a single malware APK, and the color indicates its assigned class: Adware (1), Trojan (2), Riskware (3), or Others (4). Notably, this visualization excludes benign applications, thereby focusing exclusively on malicious behaviors.

We observe that Trojan (2) and Adware (1) samples dominate the central dense regions of the UMAP projection, indicating that many features are shared between these families. Both often rely on similar mechanisms for persistence, user deception, and data exfiltration, which likely contributes to their overlap.

Riskware (3) samples exhibit a more dispersed distribution; however, small localized clusters of this category can still be observed. This dispersion may reflect the heterogeneous nature of Riskware, where the boundary between risky but not inherently malicious behavior and clearly malicious activity becomes blurred. The widespread nature also suggests that Riskware often adopts techniques or feature signatures from other families.

The Others (4) category appears mainly at the peripheries and intermixes within the central clusters. This distribution is expected, as the ``Others'' class likely includes outliers, mixed-behavior samples, or malware exhibiting multi-family characteristics.

\section*{Usage Notes}



For the implementation, we employed Python (versions 3.8 and 3.10) and Bash scripting (version 5.0.17). 
We used the AndroGuard library (version 3.3.5) to implement the extraction module and NetworkX (version 2.2) to construct the API call graph. 
To manipulate JSON and CSV files, we adopted the Python Pandas library (version 1.3.5). 
We also utilized other Python modules, including Lxml (version 4.5.0), Numpy (version 1.22.3), and Termcolor (version 2.3.0), for data handling, processing, and highlighting.
For model training, we used a machine equipped with an AMD Ryzen 7 5800X processor, an NVIDIA GeForce RTX 3090 (24GB GDDR6X), 64 GB of DDR4 3200 MHz RAM, and a 512 GB M.2 NVMe SSD.

\section*{Code availability}


The dataset MH-1M is publicly available and can be accessed through the Figshare \cite{braganca2025} and GitHub \cite{hendrio2024mh1mGitHub} repositories. The MH-1M raw data can be found in compressed format in the Harvard Dataverse repository \cite{Braganca2025MH1M}.
The MH‑1M dataset was constructed using a pipeline powered by two open-source tools developed by \cite{rocha2023am_generator_explorer}.
AMGenerator \cite{malwarehunter2023amgenerator} is an improved version of ADBuilder \cite{vilanova2022adbuilder} that handles the Android APK acquisition, static feature extraction (e.g., permissions, intents, API calls, opcodes), and VirusTotal-based malware labeling. It uses the VirusTotal API to retrieve up-to-date threat assessments across millions of samples. 
AMExplorer \cite{malwarehunter2023amexplorer} processes and consolidates the outputs from AMGenerator, integrating features and metadata into a unified, structured dataset optimized for analysis and machine learning tasks.



\bibliographystyle{IEEEtran}
\bibliography{sample}


\section*{Acknowledgements} 


This research was partially funded, as provided for in Arts. 21 and 22 of Decree No. 10.521/2020, under Federal Law No. 8.387/1991, through agreement No. 003/2021, signed between ICOMP/UFAM, Flextronics da Amazônia Ltda., and Motorola Mobility Comércio de Produtos Eletrônicos Ltda. 
This work was financed in part by the Coordenação de Aperfeiçoamento de Pessoal de Nível Superior - Brasil (CAPES-PROEX) - Finance Code 001,   partially supported by the Amazonas State Research Support Foundation - FAPEAM -  through the POSGRAD project 2024/2025, and partially supported by the Research Support Foundation of the State of Rio Grande do Sul (FAPERGS) under grant agreements 24/2551-0001368-7 and 24/2551-0000726-1.

\section*{Author contributions statement}

Conceptualization, H.B; 
methodology, H.B., V.R., J.A., D.K., E.F.; 
software, H.B., V.R., J.A.; 
validation, H.B., V.R., J.A., D.K., E.F.; 
formal analysis, H.B., V.R., J.A., D.K., E.F.; 
investigation, H.B., V.R., J.A.; 
resources, H.B., V.R., J.A.; 
data curation, H.B., V.R., J.A.; 
writing---original draft preparation, H.B, D.K., E.F.; 
writing---review and editing, H.B., V.R., J.A., D.K., E.F.;  
visualization, H.B., D.K., E.F.; 
supervision, D.K., E.F.; 
project administration, D.K., E.F.; 
funding acquisition, D.K., E.F.;


\section*{Competing interests} 

The authors declare no competing interests. The financial support received for this project had no influence on the study’s design, data collection, analysis, interpretation, or the decision to publish.

\end{document}